\documentclass[12pt]{iopart}
\usepackage{cite}
\usepackage{graphicx}
\graphicspath{{./plots/}}
\usepackage{iopams}
\usepackage{mathptmx}
\usepackage{textcomp}
\usepackage{mymacros}

\renewcommand{\d}{\mathrm{d}}
\renewcommand{\i}{\mathrm{i}}
\renewcommand{\Re}{\mathop{\mathrm{Re}}}
\renewcommand{\up}{{\uparrow}}
\renewcommand{\down}{{\downarrow}}
\newcommand{\void}[1]{}

\begin{document}

\title{Landau-Zener tunnelling in dissipative circuit QED}

\author{David Zueco, Peter H{\"a}nggi and Sigmund Kohler}
\address{Institut f{\"u}r Physik,
	Universit{\"a}t Augsburg, Universit{\"a}tsstra{\ss}e 1,
	86135 Augsburg, Germany}
\ead{\mailto{david.zueco@physik.uni-augsburg.de}}

\date{\today}

\begin{abstract}
We investigate the influence of temperature and dissipation on the
Landau-Zener transition probability in circuit QED.  Dissipation is
modelled by coupling the transmission line to a bath of harmonic
oscillators, and the reduced density operator is
treated within Bloch-Redfield theory.  A phase-space representation
allows an efficient numerical implementation of the resulting master
equation.  It provides reliable results which are valid even for
rather low temperatures.  We find that the spin-flip probability as a
function of temperature and dissipation strength exhibits a
non-monotonic behaviour.  Our numerical results are complemented by
analytical solutions for zero temperature and for vanishing
dissipation strength.
\\[2ex] \today
\end{abstract}


\section{Introduction}
\label{sec:intro}

The demonstration of coherent quantum dynamics in superconducting flux
and charge qubits \cite{Nakamura1999a, Vion2002a, Chiorescu2003a}
represents a major step towards a solid-state implementation of a
quantum computer.
Manipulation and readout of a qubit can be achieved by a controlled
interaction with an electromagnetic circuit.  For charge qubits
implemented with Cooper-pair boxes, this can be established either by
a capacitive coupling to a transmission line \cite{Wallraff2004a,
Blais2004a} or an oscillating circuit \cite{Grajcar2004a,
Sillanpaa2005a} which can be modelled as a harmonic oscillator.  The
corresponding qubit-oscillator model also plays a role for the
description of a flux qubit that couples inductively to an embracing
dc SQUID \cite{Chiorescu2004a}.  These setups represent solid-state
realizations of a two-level atom in an optical cavity
\cite{Raimond2001a, Walther2006a}.
As compared to optical realizations, most solid-state
implementations are characterised by a much larger ratio between
qubit-oscillator coupling and oscillator linewidth
\cite{Blais2004a}.

Circuit QED experiments have already demonstrated quantum coherent
dynamics \cite{Sillanpaa2006a}, measurements with low backaction
\cite{Sillanpaa2005a, Duty2005a}, and the creation of entanglement
between two qubits in a cavity \cite{Sillanpaa2007a, Majer2007a}.
A further crucial prerequisite for a working quantum computer
is quantum state preparation, i.e.\ the initialisation of the qubits.
It has been suggested to achieve this goal by switching a control
parameter through an avoided crossing with an intermediate velocity,
such that the resulting Landau-Zener transition is neither in the
adiabatic nor in the diabatic limit \cite{Saito2006a, Wubs2007a}.
Then the avoided crossing effectively acts like a beam splitter.

The physics of Landau-Zener tunnelling is also of relevance for
adiabatic quantum computation which relies on the time evolution of
the ground state of a quantum system of a slowly time-dependent
Hamiltonian \cite{Farhi2001a}.  In this scheme, a relevant source of
errors are Landau-Zener transitions at avoided crossings between
adiabatic energy levels.  For an isolated two-level system, the
corresponding transition probability has been derived in the classic
works by Landau, Zener, and Stueckelberg \cite{Landau1932a,
Zener1932a, Stueckelberg1932a}.  When the qubit is
coupled to a heat bath, this probability may change significantly.
However, there exist also system-bath interactions for which the
transition probability at zero temperature surprisingly is not
affected at all by the coupling to a harmonic-oscillator bath
\cite{Wubs2006a} or a spin bath \cite{Saito2007a, Wan2007a}. For a
heat bath with Ohmic spectral density, the transition probability at
high temperatures is bath-independent as well \cite{Gefen1987a,
Shimshoni1991a, Ao1989a, Ao1991a, Pokrovsky2007a}.  The same holds
true for the coupling to a classical noise source \cite{Kayanuma1984a,
Saito2002a, Vestgarden2008a}.

With this work we study finite temperature Landau-Zener transitions of
a qubit that couples via a harmonic oscillator to a heat bath.  For
the qubit itself, this represents a  case of a structured heat bath,
because owing to its linearity, the oscillator plus the bath can be
considered as an effective bath with a peaked spectral density
\cite{Garg1985a, Ford1988a, Thorwart2000a, Thorwart2004a,
Wilhelm2004a, VanKampen2004a, Ambegaokar2006a, Ambegaokar2007a,
Nesi2007a} for which we provide results for the dissipative
Landau-Zener problem for finite
temperatures.  In doing so, we restrain from eliminating this extra
oscillator degree of freedom because in the present context, the
oscillator dynamics is of experimental interest as well.
The organisation of the work is as follows. In
section~\ref{sec:model+master}, we introduce the qubit-oscillator-bath
model and its treatment within Bloch-Redfield theory.
Section~\ref{sec:thermal} is devoted to Landau-Zener transitions for a
thermal initial state, while in section~\ref{sec:diss}, we investigate
dissipative transitions.  Details of the derivation of the quantum
master equation and its numerical treatment are deferred to the
appendices.

\section{Model and master equation}
\label{sec:model+master}

\subsection{Landau-Zener dynamics in circuit QED}
\label{sec:system}

Circuit QED involves a Cooper pair box that couples capacitively to a
transmission line which is described as a harmonic oscillator
\cite{Blais2004a}.  The Cooper pair box is formed by a dc SQUID such
that the effective Josephson energy $\EJ = E_\mathrm{J}^0
\cos(\pi\Phi/\Phi_0)$ can be tuned via an external flux $\Phi$, where
$\Phi_0$ denotes the flux quantum.  We assume that $\Phi$ can be
switched such that $\EJ= \hbar vt$, $v>0$, in a sufficiently large time
interval.  The capacitive energy $\frac{1}{2}\EC(N-N_g)^2$ is
determined by the number $N$ of Cooper pairs on the island and the
scaled gate voltage $N_g$.
In the charging limit $\EC\gg\EJ$, only the two states $|N=0\rangle$
and $|N=1\rangle$ determine the physics, and one defines the qubits states
$|\up\rangle \propto |0\rangle+|1\rangle$ and $|\down\rangle \propto
|0\rangle-|1\rangle$.  Then, at the charge degeneracy point $N_g=1/2$,
the Hamiltonian reads
\begin{equation}
\label{Rabi}
\Hs =
     -\hbar \frac{vt}{2}\sigma_z
     +\hbar \g\sigma_{x} ( a^{\dag} + a )
     +\hbar\Omega a^{\dag} a ,
\end{equation}
where the first term describes the qubit in pseudo spin notation with
$\sigma_z|\up,\down\rangle = \pm|\up,\down\rangle$.
The second term refers to the coupling of the qubit to the fundamental
mode of the transmission line which is described as a harmonic
oscillator with the usual bosonic creation and annihilation operators
$a^\dag$ and $a$ and the energy eigenstates $|n\rangle$,
$n=0,1,\ldots,\infty$.
Note that below we also consider the case of strong qubit-oscillator
coupling for which a rotating-wave approximation for the coupling
Hamiltonian is not justified.

\begin{figure}[tb]
\centerline{\includegraphics{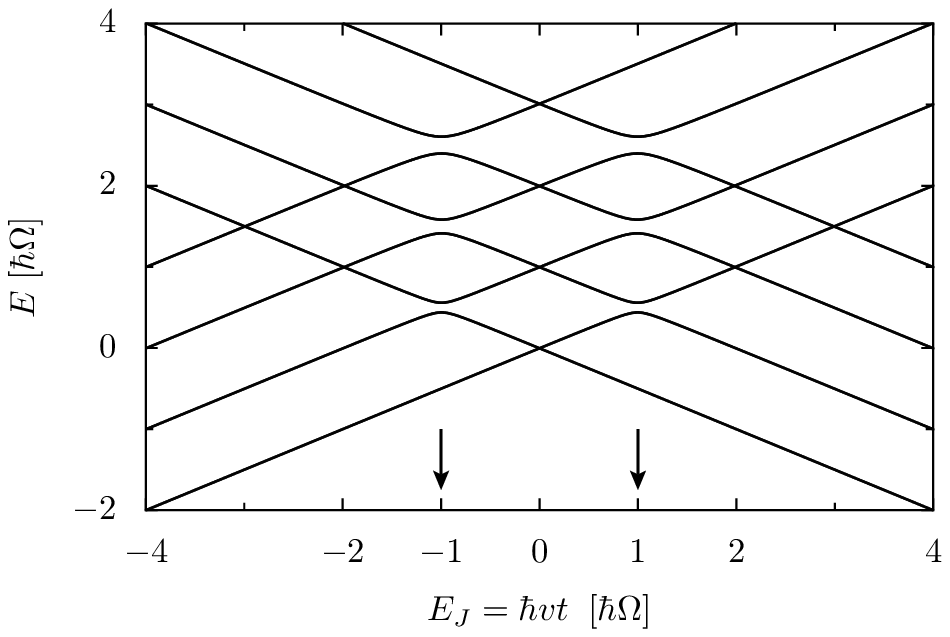}}
\caption{\label{fig:espectrum} Adiabatic energy levels of the
  qubit-oscillator Hamiltonian \eref{Rabi} as a function of
  the Josephson energy which is swept at constant velocity such that
  $\EJ= \hbar vt$.  The arrows mark the values of $\EJ$ where the
  anticrossings are located.}
\end{figure}%
When the effective Josephson energy is switched from a large negative
value to a large positive value, the system exhibits interesting
quantum dynamics which can be qualitatively understood by computing
the adiabatic energies as a function of time, see
figure~\ref{fig:espectrum}: For low temperatures, we can assume that
both the qubit and the oscillator are initially in their instantaneous
ground state $|\up,0\rangle$.  As time evolves, the system will
adiabatically follow the state $|\up,0\rangle$  until at
time $t=\Omega/v$, an avoided crossing is reached.  Then the
system will evolve into the superposition $\alpha(v)|\up,0\rangle +
\beta(v) |\down,1\rangle$ with velocity-dependent probability
amplitudes.  This means that by adjusting the sweep velocity $v$, one
can generate a single-photon state or qubit-oscillator entanglement
\cite{Saito2006a}.
For the time-evolution from $t=-\infty$ to $t=\infty$, the
corresponding bit-flip probability can be evaluated exactly and reads
\cite{Saito2006a}
\begin{equation}
\label{PLZ}
\Pud = 1 - \e^{-2 \pi g^2/ v} .
\end{equation}
Note that this generalisation of the Landau-Zener formula is also
valid for large qubit-oscillator coupling $g\gg\Omega$ for which
more than two levels are relevant and, thus, the scenario sketched above
becomes more involved.

\subsection{Dissipative dynamics}
\label{sec:qme}

Dissipative effects in an electromagnetic circuit are characterised by
an impedance $Z(\omega)$ which, within a quantum mechanical description,
can be modelled by coupling the circuit bi-linearly to its
electromagnetic environment \cite{Yurke1984a}.  This provides the
system-bath Hamiltonian \cite{Caldeira1983a, Hanggi1990a, Weiss1999a,Hanggi2005a}
\begin{equation}
\label{Htot}
\Htot = \widetilde{\Hs}
+ (a^{\dag} + a) \sum_{k} \hbar\ck (b^{\dag}_{k} + b_{k})
+ \sum_k \hbar\omega_k b_k^\dag b_k,
\end{equation}
where $\widetilde{\Hs} = \Hs + (a^{\dag} + a)^{2} \sum
\hbar \ck^{2}/2\omega_{k}$ is the Hamiltonian of the qubit and the
transmission line augmented by a counterterm, such that a frequency
renormalisation due to the coupling to the bath is cancelled
\cite{Caldeira1983a, Leggett1987a, Hanggi1990a, Hanggi2005a}.
The second and third term describe the capacitive coupling of the
transmission line to a bath of harmonic oscillators.  The bath is
fully characterised by its spectral density
\begin{equation}
\label{JZ}
J(\omega)
=
\sum_k \ck^{2} \, \delta (\omega_{k}-\omega)
=
\frac{1}{2 \pi} \sqrt{\frac {L}{C}} \;
\omega \Re Z(\omega) \; .
\end{equation}
The second equality relates the system-bath model to classical circuit
theory, where $L$ and $C$ are the specific inductance and capacitance
of an effective transmission line that forms the dissipative
environment.  This relation can be established by comparing the
resulting quantum Langevin equation with Kirchhoff's laws
\cite{Yurke1984a, Ingold1992a}.

Since we are only interested in the dynamics of the qubit and the
oscillator, all relevant information is contained in the reduced
density operator $\dm = \tr_\mathrm{B}\dm_\mathrm{total}$ which is
obtained by tracing out the bath degrees of freedom.  For weak
system-bath coupling, the bath can be eliminated within Bloch-Redfield
theory \cite{Redfield1957a, Blum1996a} in the following way:
Assuming that initially the bath is at thermal equilibrium and not
correlated with the system, $\dm_\mathrm{tot} \propto \dm \otimes\exp(-\sum_k
\hbar\omega_k b_k^\dag b_k/k_\mathrm{B}T)$, one derives within
perturbation theory the quantum master equation
\begin{equation}
\label{BR}
\fl
\frac{\d}{\d t}\dm
=
-\frac{\iu}{\hbar} [\Hs, \dm]
-
\int_{0}^{\infty} \diff \tau \,
\Big \{
\Sym (\tau)[\X,[\X(-\tau), \dm]]
+
\iu
\Ant(\tau)
[\X,[\X(-\tau),\dm]_{+}]
\Big \}
\; ,
\end{equation}
where $[A,B]_{+}= AB + BA$ and $[A,B]= AB - BA$ denote the
anti-commutator and the commutator, respectively, while the scaled
``position'' $Q=a+ a^{\dag}$ is the system operator to which the bath
couples.
The first term describes the unitary evolution generated by the system
Hamiltonian, while the second one captures the dissipative influence
of the environment.  The dissipative terms depend on the system
dynamics through the Heisenberg operator $\X(t) = \e^{\iu \Hs
  t/\hbar} \X \e^{-\iu \Hs t /\hbar}$.  The bath enters via the
symmetric and the antisymmetric correlation functions
\begin{eqnarray}
\label{corr-sym}
\Sym (t)= \frac{1}{2}\langle [\B(t), \B]_{+} \rangle_{\rm eq}
&=&
\int_{0}^{\infty}
\diff \omega
\;
J(\omega)
\coth \Big( \frac{\hbar\omega}{2k_\mathrm{B}T}\Big)
\cos(\omega t)
\\
\label{corr-ant}
\Ant(t) =\frac{1}{2\iu}\langle [\B(t), \B] \rangle_{\rm eq}
&=&
\int_{0}^{\infty}
\diff \omega \; J(\omega) \sin(\omega t)
\end{eqnarray}
of the effective bath operator $\B=\sum_k c_k (b_k^\dag + b_k)$.
The Markovian master equation implies that the bath stays always close
to equilibrium and that no system-bath correlations build up.

For an explicit form of the master equation, we still need to evaluate
the Heisenberg time evolution of the system operator $\X(t)$.
For a time-independent system Hamiltonian, this can be done exactly by
a decomposition into the energy eigenbasis.  Then one obtains the
standard Bloch-Redfield approach \cite{Redfield1957a, Blum1996a}. For
periodically driven systems, the coherent dynamics is solved by the
Floquet states which provide an appropriate basis \cite{Kohler1997a}.
The Hamiltonian \eref{Rabi}, however, possesses a more general
time-dependence and, thus, we have to resort to further
approximations.
If $\EJ \sim \hbar\Omega$, one could employ a rotating-wave
approximation (RWA) for the cavity-qubit coupling, $\sigma_{x} (a +
a^{\dag}) \cong \sigma_{+} a + \sigma_{-} a^{\dag}$ with $\sigma_{\pm}
= \sigma_{x} \pm \iu \sigma_{y}$.
Within RWA, the Hamiltonian \eref{Rabi} turns into the
Jaynes-Cummings model for which the eigenvalues and eigenvectors are
known \cite{Gardiner1991a}.
Far off resonance, by contrast, i.e.\ for $\hbar\Omega \gg \EJ$ or
$\hbar\Omega \ll \EJ$, an adiabatic approximation for either the qubit
or the harmonic oscillator \cite{Tornberg2007a, Irish2005a} is helpful.
For the present case of a Landau-Zener sweep, however, the Josephson
energy assume all values from $-\infty$ to $+\infty$, such that
generally none of these approximations is well suited.
Therefore, we resort to a weak-coupling approximation in the
qubit-oscillator interaction $g$.  The corresponding solution of the
Heisenberg equations for the dimensionless position operator $\X(t)$
is derived in the \ref{app:qme} and reads
\begin{equation}
\X(t) = a^{\dag} \e^{\iu \Omega t} + a \e^{-\iu \Omega t}
        + \g \big[ \Ic(t) \sigma_{x} - \Is (t) \sigma_{y} \big] ,
\label{Xt}
\end{equation}
with the time-dependent functions
$
\Ic (t) = \Omega [ \cos(\Omega t) - \cos(\omega_\mathrm{J}t) ]
/(\Omega^{2} - \omega_\mathrm{J}^{2})
$
and
$
\Is (t) = [ \omega_\mathrm{J} \sin(\Omega t) - \Omega\cos
(\omega_\mathrm{J} t) ]/(\Omega ^{2} - \omega_\mathrm{J} ^{2})
$,
where $\omega_\mathrm{J} = \EJ/\hbar$.
In the derivation of this expression, we have neglected all terms of
order $(\g t)^{2}$.
Note that this approximation is  used only for the evaluation of the
dissipative kernels of the master equation~\eref{BR}, while for the
solution of the master equation, the qubit-oscillator coupling is
treated exactly.
This means that we  neglect dissipative terms of the order
$ \dmp \g^{2} / \Omega^3 $ only,
which is justified for a weak oscillator-bath coupling.

For the further evaluation, we still need to specify the spectral
density $J(\omega)$.  Assuming that the environment is strictly
ohmic and recalling that $\Omega = 1 /\sqrt{LC}$, we obtain from
Eq.~\eref{JZ}
\begin{equation}
\label{Jw}
J(\omega) = \frac{\dmp}{2 \pi \Omega} \omega
\end{equation}
with the effective damping rate $\dmp$.  Inserting expressions
\eref{corr-sym}--\eref{Xt} into Eq.~\eref{BR}, we arrive at the
explicit quantum master equation (see again \ref{app:qme})
\begin{eqnarray}
\nonumber
\frac{\d}{\d t} \dm
&=&
-\frac{\iu}{\hbar} [\Hs, \dm]
-\iu \frac{\gamma}{4 \Omega} [\X,[\dot \X,\dm]_{+}]
- \frac{\gamma}{4} \Dpp [\X,[\X,\dm]]
\\ &&
- \frac{\gamma}{4 \Omega} \Dxp [\X,[\dot \X,\dm]]
- \frac{\gamma}{4} \Dsigma [\X, [\sigma_{z}, \dm]] ,
\label{QME-cqed}
\end{eqnarray}
with the operator $\dot \X = \iu/\hbar[\Hs, \X] = \iu \Omega (a^{\dag}- a)$
and diffusion constants
\begin{eqnarray}
\Dpp &\equiv &
\coth \Big ( \frac{\hbar \Omega}{2k_\mathrm{B}T} \Big )
\; ,
\\
\Dxp
&=&
\frac{\nu_{1}  \cutoff^{2}}{2(\cutoff^{2} + \Omega^{2})}
\sum_{n=-\infty}^{\infty}
\frac{\Omega^{2} - \nu_{n} \cutoff}
{(\nu_{n}+ \cutoff)(\nu_{n}^{2}+ \Omega^{2})} ,
\end{eqnarray}
which refer to momentum diffusion and to cross-diffusion, respectively,
while $\nu_{n} = 2\pi\kB T/\hbar$ denotes the Matsubara frequencies and
$\cutoff$ is a high frequency Drude cutoff.
The prefactor of the last term contains an effective force
\begin{equation}
\label{Fsigma}
\Dsigma =
\frac{\g}{2(\Omega^{2} - \omega_\mathrm{J}^{2})}
\Big [
\omega_\mathrm{J} \coth\Big(\frac{\hbar\omega_\mathrm{J}}{2k_\mathrm{B}T}\Big)
- \Omega \coth \Big( \frac{\hbar\Omega}{2k_\mathrm{B}T} \Big)
\Big ]
\end{equation}
which acts on the oscillator and depends on the state of the qubit.
This means that the qubit dynamics influences the dissipative terms
despite the fact that it couples to the bath only indirectly via the
oscillator.  This influence vanishes in the high-temperature limit,
where in addition the diffusion coefficients become $\Dpp=2
k_\mathrm{B}T/\hbar \Omega$ and $\Dxp=0$, such that dissipative terms
in Eq.~\eref{Jw} reduces to those of the well-known form derived in
Ref.~\cite{Caldeira1983b}.

The cross-diffusion diffusion term $\propto\Dxp$ can be rather cumbersome
due to its explicit dependence on the cutoff $\cutoff$, which yields an
ultraviolet divergence.  Still it is possible to avoid its explicit
evaluation and, thus, to render the cutoff obsolete with
renormalisation arguments:
The divergence and its regularization can be related to the
ultraviolet divergence for the equilibrium momentum variance $\langle
\dot{\X}^{2} \rangle_\mathrm{eq}$ for which we find the relation
\begin{equation}
\frac{\gamma}{\Omega} \Dxp
=
\langle \X^{2} \rangle_{{\rm eq}}
-
\Omega^{2} \langle \dot{\X}^{2} \rangle_{{\rm eq}} ,
\end{equation}
where $\langle \,\cdots \, \rangle_{{\rm eq}}$ denotes the thermal
average.
It turns out that in the solution of the quantum master equation,
$\Dxp$ appears only in the combination $\langle \X^{2} \rangle_{{\rm eq}} =
\dmp \Dxp/\Omega + \Dpp$ (see Eq.~\eref{transfdos}), neglecting $\Dxp$
is consistent with a weak-coupling approximation.
Since for the harmonic oscillator the exact relation $\langle \X^{2}
\rangle_{{\rm eq}} = \Dpp/\Omega^{2}$ holds \cite{Weiss1993a}, the
replacement $\Dxp\to 0$ provides the correct equilibrium
expectation values, even in the limit of low temperatures.


\subsection{Solving the master equation in phase space}

The numerical solution of the quantum master equation~\eref{QME-cqed}
requires an appropriate basis expansion.  For the qubit, we choose the
eigenstates of $\sigma_z$, i.e.\ $|{\uparrow}\rangle$ and
$|{\downarrow}\rangle$.  The resulting matrix elements $\rho_{ij}$
with $i,j=\up,\down$ are operators in the Hilbert space of the harmonic
oscillator.  For these, at first sight, the natural basis is provided
by the Fock states $|n\rangle$.  With increasing temperature,
however, good convergence is obtained only for a relatively large
number of states.  Then it is advantageous to transform the operators
$\rho_{ij}$ to a phase-space quasi distribution like the Wigner
representation \cite{Hillery1984a} which can be defined as
\begin{equation}
W_{jj'}(x,p)
= \frac{1}{2\pi} \int_{- \infty}^{\infty} \diff y \exponent {\iu y p}
\rho_{jj'}(x -\frac{1}{2} y, x+\frac{1}{2} y) \; .
\end{equation}
The actual transformation can be accomplished by using Bopp operators
\cite{Hillery1984a}.
Introducing for the qubit part the matrix notation
\begin{eqnarray}
\Wm(x,p)
\equiv
\left ( \begin{array}{cc} W_{\up \up} & W_{\up \down}
                          \\
                          W_{\down \up} & W_{\down \down}
\end{array} \right) ,
\end{eqnarray}
one obtains for the master equation \eref{QME-cqed} the Wigner
representation
\begin{equation}
\partial_{t} \Wm
=
\Lho \Wm
+ \iu \frac{\EJ}{2 \hbar} [\sigma_{z}, \Wm]
+ \g \partial_{p} [\sigma_{x}, \Wm]_{+}
+ \iu  \g x [\sigma_{x}, \Wm]
- \iu \gamma  \Dsigma  \partial_{p} [\sigma_{x}, \Wm] .
\label{QME-cqed-ps}
\end{equation}
The Fokker-Planck-like operator
\begin{equation}
\label{Lho}
\Lho
\equiv
- \Omega \big ( p \partial_{x} + x \partial_{p} \big )
+ \gamma \partial_{p} p
+ \gamma \Dpp \partial^{2}_{p}
+ \dmp \Dxp \partial^{2}_{xp}
\end{equation}
governs the dissipative time-evolution of the harmonic oscillator,
while the next three terms describe the coherent time evolution of the
qubit and its coupling to the oscillator.  The last term refers to the
modification of the dissipative terms that stems from the
qubit-oscillator coupling.

The main advantage of this representation comes from the fact that the
oscillator part $\Lho$ is formally identical with the Klein-Kramers
equation \cite{Risken1989a,Hanggi1990a} for the classical dissipative oscillator,
which allows one to adapt techniques for solving Fokker-Planck
equations to quantum master equations \cite{Kohler1997a,
GarciaPalacios2004a, Coffey2007a}.
In particular, we will use the eigenfunctions $\phi_{nn'}(x,p)$ of
$\Lho$ which obey the eigenvalue equation \cite{Titulaer1978a}
\begin{equation}
\Lho\phi_{nn'}(x,p) = (n\lambda+n^*\lambda^*)\phi_{nn'}(x,p) ,
\qquad n,n'=0,1,2,\ldots,
\end{equation}
where $\lambda = -\gamma/2+\i (\Omega^2-\gamma^2/4)^{1/2}$; see
\ref{app:Lho}.  The ``ground state'' $\phi_{00}$ is the Wigner
representation of the density operator of the harmonic oscillator at
thermal equilibrium.  Thus, if the oscillator stays close to
equilibrium, the decomposition of the density operator can be
performed with only a few basis states --- irrespective of the
temperature.  The resulting equations of motion for the coefficients
can be found in \ref{app:be}.

\section{Landau-Zener tunnelling at finite temperature}
\label{sec:thermal}

Thermal effects can modify the Landau-Zener transition
probability~\eref{PLZ} even in the absence of dissipation, i.e.\ for
$\gamma=0$.  Then the natural initial state is no longer the (initial)
ground state $|\up,0\rangle$, but rather the canonical ensemble
\begin{equation}
\label{CIT}
\dm (t=-\infty)
= \ket \up \bra \up \otimes \frac{\e^{-\Hho/\kB T}}{\Z} ,
\end{equation}
where $\Hho = \hbar a^{\dag} a$ is the cavity Hamiltonian and $\Z =
\tr[\exp({-\Hho/\kB T})]$ the partition function at temperature $T$.
Note that initially the qubit splitting is infinite, so that even at
finite temperature, only the qubit ground state is populated.

\begin{figure}
\includegraphics[width=4.2cm]{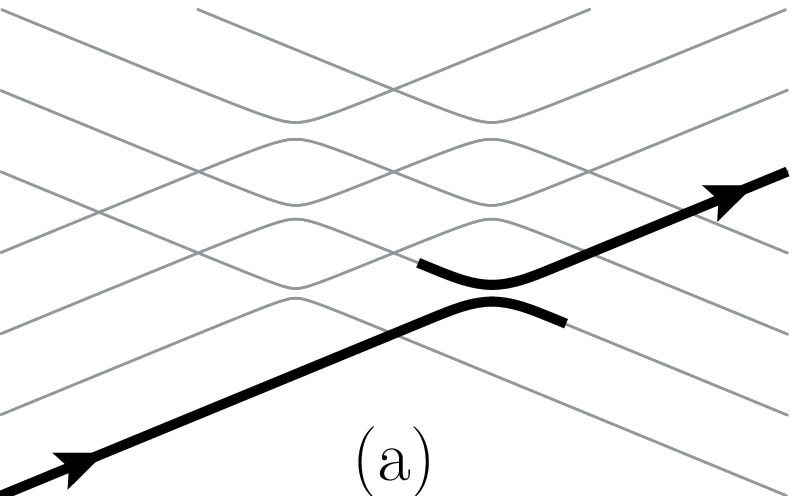}\hfill
\includegraphics[width=4.2cm]{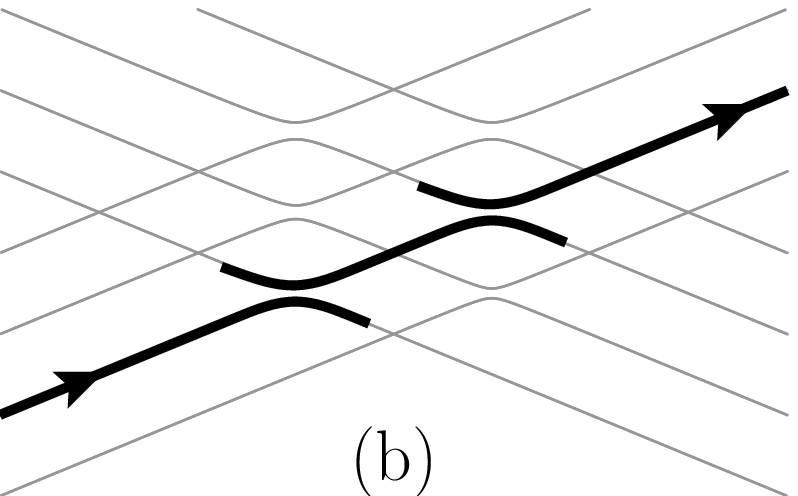}\hfill
\includegraphics[width=4.2cm]{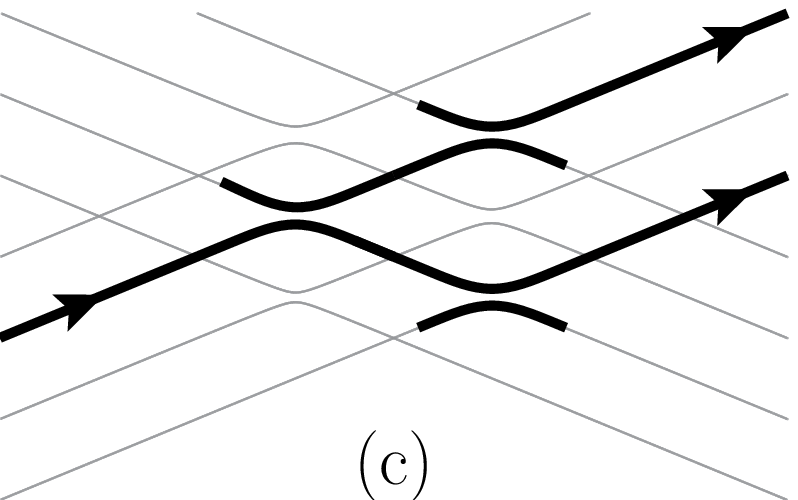}
\caption{Transition paths from the initial states
$|\up,n\rangle$, $n=0,1,2$, to the qubit state $|\up\rangle$, i.e.\ those
contributing to the probability $P_{\up \to \up}$ in the
individual-crossing approximation, which is valid for weak qubit-oscillator
coupling~$g\ll\Omega$.}
\label{fig:paths}
\end{figure}

\begin{figure}
\centerline{\includegraphics{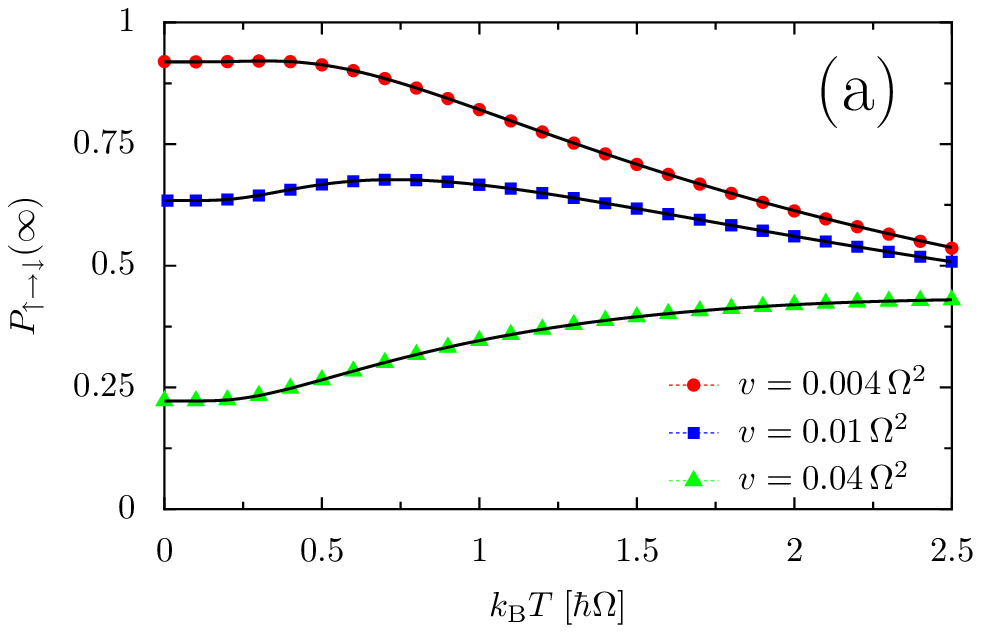}
\includegraphics{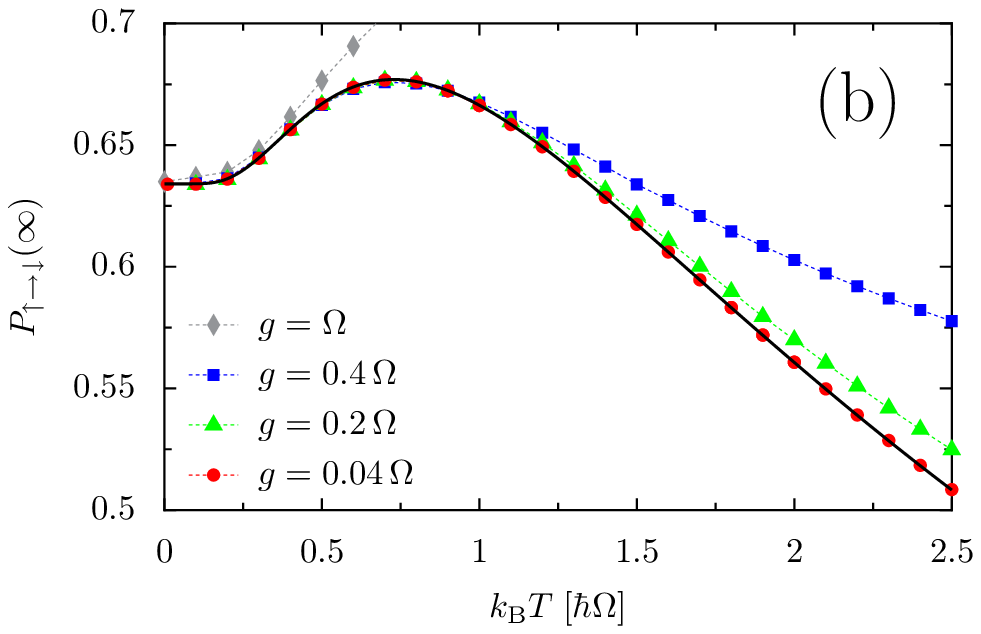}}
\caption{Spin-flip probability $\Pud$ as a function of the temperature
for various sweep velocities and $\g=0.04\Omega$ (a) and for various
qubit-oscillator coupling strengths and $v=0.01\Omega^2$ (b).
The symbols mark the numerical solution of the Liouville-von Neumann
equation (Eq.~(\ref{QME-cqed-ps}) for $\gamma=0$), while the solid lines
refer to the individual-crossing result \eref{Puuana}.  The dashed
lines are a guide to the eye.
}
\label{fig:Tdep_nodiss}
\end{figure}

Since $\dmp=0$, the quantum master equation~\eref{QME-cqed} reduces to
the unitary von Neumann equation for the Hamiltonian \eref{Rabi},
which is independent of the approximations made in the derivation of
the master equation.  Then the transition probability $\Puu$ is the
thermal average of the transition probabilities for the initial states
$\ket{\up,n}$, i.e.
\begin{equation}
\label{Puu_gral}
\Puu
= \sum_{n} p_{n} \; P_{\uparrow, n \to \uparrow}
= \sum_{n,m} p_{n} \; P_{\uparrow, n \to \uparrow, m} ,
\end{equation}
where $p_{n} ={\e^{-n\hbar\Omega / \kB T}}/\Z$ and
$P_{\uparrow, n  \to \uparrow, m} = \vert \bra {\up ,  n } U (\infty;
0) \ket {\up ,  m} \vert ^{2}$.
It is worth noting that only terms with $n > m$ contribute to the sum,
due to the ``no-go-up'' theorem \cite{Saito2007a} which states that
$P_{\up,n\to\up,m}=0$ for $m>n$.  For the computation of the remaining
probabilities $P_{\up, n \to \up, m}$ we need to resort to an
approximation: If all avoided crossings in the adiabatic qubit-oscillator
spectrum (Fig.~\ref{fig:espectrum}) are well separated, one can treat
the transitions at the avoided crossing as being independent of each
other and compute the transition probabilities $P_{\up,n\to\up,m}$ as
joint probabilities \cite{Wan2007a}.
In the vicinity of an avoided crossing, the qubit-oscillator
Hamiltonian is described by the two-level system $H_\mathrm{TLS} =
\frac{1}{2}vt\sigma_z+\frac{1}{2}\Delta\sigma_x$ with sweep velocity
$v$ and some level splitting $\Delta$.  The corresponding probability
for a non-adiabatic transition is given by the standard Landau-Zener
expression \cite{Landau1932a, Zener1932a, Stueckelberg1932a}
\begin{equation}
\label{standardPLZ}
w(\Delta) = \e^{-\pi\Delta^2/2v} .
\end{equation}
For the Hamiltonian~\eref{Rabi}, the avoided crossings are formed only
between states $|\up,n\rangle$ and $|\down,n+1\rangle$ at
times $t=\mp\Omega/v$ with the level splitting
\begin{equation}
\label{Delta}
\Delta_n = 2 \g \sqrt{n+1}  .
\end{equation}
Then the only paths that connect two qubit states $|\up\rangle$ are
sketched in Fig.~\ref{fig:paths}.  For the initial states
$|\up,0\rangle$ and $|\up,1\rangle$, the probabilities to end up in
the qubit state $|\up\rangle$ then become $P_{\up,0\to\up}=w(2g)$ and
$P_{\up,1\to\up}=w(2g)w(2g\sqrt{2})$, respectively.  For oscillator
states with $n>1$, two final oscillator states are possible.
Assuming that interference terms do not play any role, we find the
transition probability
\begin{equation}
\label{Pn}
P_{\up,n\to\up} = w(2g\sqrt{n}) w(2g\sqrt{n+1})
+ [1-w(2g\sqrt{n})] [1-w(2g\sqrt{n-1})] ,
\end{equation}
which formally also holds for $n=0,1$.
Inserting \eref{Pn} into \eref{Puu_gral}, we obtain the transition
probability $\Pud = 1- \sum_n p_n P_{\up,n\to\up}$, where the sum can
be identified as a geometric series.  Evaluating this series, yields
a central result, namely the Landau-Zener transition probability for
finite temperature and weak qubit-oscillator coupling:
\begin{equation}
\Pud
=
\Bd (g^{2}/v) - \Bd (2 g^{2} /v ) \, \e ^{-2\pi\g^2/\vel}
+ \big [ \Bd (g^{2} /v ) - \Bd (2 g^{2} /v) \big ] \e^{2 \pi \g^2/\vel}\;,
\label{Puuana}
\end{equation}
where the temperature dependence is captured by the function
\begin{equation}
\Bd (x) = \frac{1 -\e^{-\hbar \Omega/\kB T}}{1- \e^{- (\hbar
    \Omega/\kB T + 2\pi x)}} \; ,
\end{equation}
which for $x>0$ vanishes in the high-temperature limit, while $\Bd(x)
=1$ for zero temperature.  In the latter limit, expression
\eref{Puuana} becomes identical to equation~\eref{PLZ}.
The independent-crossing approximation is valid whenever the
time between the anti-crossings, $ t = 2 \vel/\Omega$, exceeds the
duration of an individual Landau-Zener transition,
$\tau_{{\rm LZ}} \sim \sqrt{1/\vel} \,
\max \{ 1, \sqrt {\Delta^{2}/ \vel} \}$ \cite{Mullen1989a}.
Inserting the explicit expression \eref{Delta} for $\Delta_n$, this
condition becomes $ \Omega > n \g $.  Thus, analytical result
(\ref{Puuana}) hold only as long as oscillator states $|n\rangle$ with
$n>\Omega/\g$ are not thermally occupied, i.e.\ for $\kB T <
\hbar\Omega^2/g$.  Fortunately, in the range of current experimental
interest, $\kB T \lesssim \hbar\Omega$ and $\g<\Omega$, this condition
is fulfilled.  The numerical results shown in figures
\ref{fig:Tdep_nodiss}(a,b) confirm the results of the individual
crossing approximation very well.  Even for higher temperatures, $\kB
T>\hbar\Omega$, we find very good agreement, provided that the
qubit-oscillator coupling $\g$ is sufficiently weak, see
figure~\ref{fig:Tdep_nodiss}(b).
Moreover, the time evolution of the population (see
figure~\ref{fig:popdyn}, below) confirms that the dynamics indeed discerns
into two individual transitions.

The temperature dependence of the spin-flip probability $\Pud$ shown
in figure~\ref{fig:Tdep_nodiss} possesses an intriguing non-monotonic
behaviour: For low temperatures, $\kB T\lesssim 0.2\hbar\Omega$, the
probability is almost temperature independent.  With an increasing
temperature, $\Pud$ first becomes larger, while it eventually
converges to zero in the high-temperature limit.  The temperature for
which the the spin-flip probability assumes a maximum is essentially
independent of the qubit-oscillator coupling $g$ and increases with
the sweep velocity.

This behaviour can be understood in the individual-crossing scenario
sketched in figure~\ref{fig:paths}.  For very low temperatures,
initially only the state $|\up,0\rangle$ is significantly populated
and, thus, only the path sketched in figure~\ref{fig:paths}(a) is
relevant.  For a temperature $T\approx\hbar\Omega/\kB$, also the
initial state $|\up,1\rangle$ becomes relevant.  For this initial
state, reaching the
final state $|\up\rangle$ requires two non-adiabatic transitions (see
figure~\ref{fig:paths}(b)) which enhances the probability to end up in
state $|\down\rangle$.  With further increasing higher temperature, states
$|\up,n\rangle$ with $n\geq 2$ start to play a role.  In the
individual-crossing picture, each of these states can evolve into four
possible finale states, two of which with spin up and two with spin
down (see figure \ref{fig:paths}(c)).  Consequently, the probability
of reaching the spin state $|\down \rangle$ is no longer enhanced.
The relevance of oscillator states with $n\geq 2$ also qualitatively
explains the fact that $\Pud$ vanishes for high temperatures: In the
adiabatic limit $v\ll\g^2$, the state $|\up,n\rangle$ evolves via
the state $|\down,n-1\rangle$ into the final state $|\up,n-2\rangle$.  In
the opposite limit of fast sweeping, the systems essentially remains
in its initial state $|n,\up\rangle$.  In both cases, the qubit
will predominantly end up in state $|\up\rangle$, which complies with
our numerical results.

The non-monotonic temperature dependence is in contrast to the
behaviour found for Landau-Zener transitions of a qubit that is
coupled to further spins, for which a monotonic temperature
dependence has been conjectured \cite{Wan2007a}.  The physical reason
for this difference is that the spin coupled to the qubit possesses
only one excited state.  Then the paths sketched in
figures~\ref{fig:paths}(b,c), which are responsible for the
non-monotonic temperature dependence, do not exist.

\section{Dissipative Landau-Zener transitions}
\label{sec:diss}

In the previous section we have studied the consequences of thermal
excitations of the initial state for the transition
probability~\eref{PLZ} in the absence of an oscillator-bath coupling,
i.e.\ for dissipation strength $\dmp=0$.  We next address the question
how dissipation and decoherence modify Landau-Zener tunnelling.

\subsection{The zero-temperature limit}
\label{sec:zeroT}

For a heat bath at zero temperature, the exact solution of the
dissipative Landau-Zener problem has been derived in recent works
\cite{Wubs2006a, Saito2007a}.  Moreover, this limit generally is
rather challenging for a master equation description of
quantum dissipation \cite{Leggett1987a, Hanggi1990a}.  Therefore the
zero-temperature limit represents a natural test bed for our
Bloch-Redfield master equation.

Let us start with a brief summary of the derivation of the exact
expression for the spin-flip probability $\Pud$, as given in
Ref.~\cite{Saito2007a}.  The central idea is to consider the
cavity plus the bath as an effective bath that consists of
``$\infty+1$'' oscillators \cite{Garg1985a, Ford1988a, VanKampen2004a,
Thorwart2004a, Ambegaokar2006a, Ambegaokar2007a, Nesi2007a}.  Then the
qubit-oscillator coupling operator $\sigma_{x}(a + a^{\dag})$ is
replaced by a qubit-bath coupling of the type $\sigma_{x} \sum_{k'}
\Xkp$, where $\Xkp$ denotes the normal coordinates of the effective
bath.  Thus the total Hamiltonian \eref{Htot} can be written in terms
of the spin-boson Hamiltonian \cite{Weiss1993a}
\begin{equation}
\label{Hsb}
{\mathcal H}_{{\rm tot}}
=
-\frac{\EJ(t)}{2} \sigma_{z}
+\hbar \g \sigma_{x} \sum_{k'} \ckp  \Xkp
+ \sum_{k'} \hbar\omegap_{k'} \bar{a}^{\dag}_{k'}\bar{a}_{k'} .
\end{equation}
Moreover, the transformation to the normal coordinates provides the
effective spectral density \cite{Garg1985a, Goorden2004a}
\begin{equation}
\label{Jeff}
J_{{\rm eff}} (\omega)
= g^2\sum_{k'} \bar c_{k'}^2 \delta(\omega-\bar\omega_{k'})
= \frac{2 \alpha \omega \Omega^{4}}
       {(\Omega^{2} - \omega^{2})^{2} + (\dmp \omega)^{2}}
\end{equation}
with the effective dissipation strength
\begin{equation}
\alpha  = \frac{4}{\pi}\frac{g^{2}}{\Omega^{3}} \dmp .
\end{equation}
For the time-dependent Josephson energy $\EJ(t) = \hbar vt$, this model
defines a dissipative Landau-Zener problem for which at $T=0$ the
spin-flip probability reads \cite{Saito2007a}
\begin{equation}
\label{Saito}
\Pud = 1 - \e ^{-2 \pi \g^2 \sum_{k'} \ckp^{2} / v} .
\end{equation}
Note that this formula is exact for any values of the coefficients
$\g$ and $ \ckp$, provided that the systems starts at $t=-\infty$ in
its ground state.
The sum $\sum_{k'} c_{k'}^{2}$ can be expressed in terms of the spectral
density \eref{Jeff}, such it becomes
\begin{equation}
\label{J-Saito}
\sum_{k'} \ckp^{2}
=
\frac{1}{g^2}\int \diff \omega  \, J_{{\rm eff}} (\omega)
=
\frac{1}{\pi} \left[
\arctan\left(
  \frac{2\Omega^{2} -\dmp^{2}}{\dmp \sqrt{4\Omega^{2}-\dmp^{2}}}
\right)
+ \frac{\pi}{2} \right] .
\end{equation}
In the limit $\gamma\to 0$, we obtain $\sum_{k'}\bar c_{k'}^2 = 1$,
such that the transition probability \eref{Saito} equals expression
\eref{PLZ} which is valid in the absence of the oscillator-bath
coupling.  For a small but finite dissipation strength $\gamma$,
$\sum_{k'}\bar c_{k'}^2 < 1$ and, thus, the spin-flip probability
becomes smaller when the oscillator is damped.

We now use this exact result as a test for the master equation
\eref{QME-cqed}.  It is worth emphasising that both problems are not
fully equivalent, because the preparations are different:
At zero temperature, the initial condition (\ref{CIT}) for the master
equation reads $\dm(t=-\infty) = \ket {\up,0} \bra {\up,0}$, i.e.\
the composed qubit-oscillator-bath system starts in the pure state
$|\up,0\rangle\otimes|0,0,\ldots\rangle$, where the latter factor
refers to the bath.
In the analytical treatment sketched in the preceding paragraph, by
contrast, the initial condition is the ground state of the total
Hamiltonian \eref{Hsb}, $|\up\rangle\otimes|\bar 0, \bar0, \bar0,
\ldots\rangle$.  Since a finite oscillator-bath coupling induces
system-bath correlations \cite{Hanggi2005a}, the oscillator is
generally entangled with the bath.
Nevertheless, in the present context, the difference should be minor,
because we consider a time evolution that starts at $t=-\infty$, such
that the oscillator-bath setting can evolve into its ground state
before Landau-Zener tunnelling occurs.

\begin{figure}
\centerline{\includegraphics{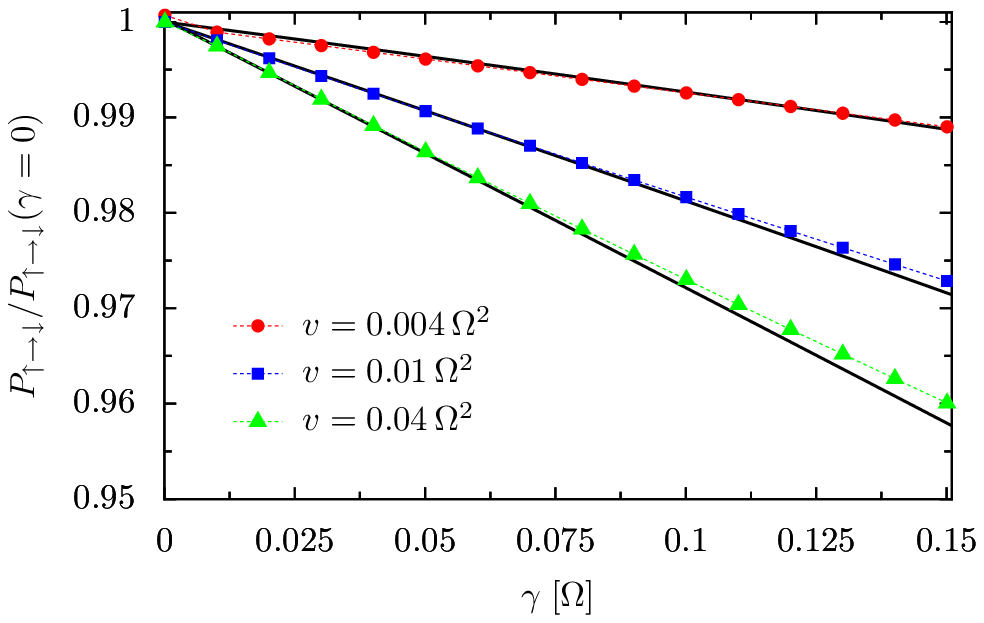}
}
\caption{Comparison of the spin-flip probability $\Pud$ at $T=0$
obtained with Bloch-Redfield theory (symbols) and the exact result,
Eqs.~\eref{Saito} and \eref{J-Saito}, as a function of the dissipation
strength for qubit-oscillator coupling strength $\g = 0.04\Omega$ and
various sweep velocities.  
The probability has been normalised to the corresponding value in the
absence of dissipation, $ \Pud (\gamma=0)$.
The deviation from the exact result is always below 1\%.
}
\label{fig:dmpdep}
\end{figure}

In figure~\ref{fig:dmpdep}, we compare the transition probabilities
obtained with the quantum master equation \eref{QME-cqed} with the
corresponding exact analytical result \eref{Saito}.  For the system
parameters used below, we find that Bloch-Redfield theory predicts
even at zero temperature the exact results with an error of less than
one percent.  Since the quality of a Markovian quantum master equation
generally  improves with increasing temperature, the
results presented below are rather reliable.

\subsection{Thermal excitations and dissipative transitions}

\begin{figure}
\centerline{\includegraphics{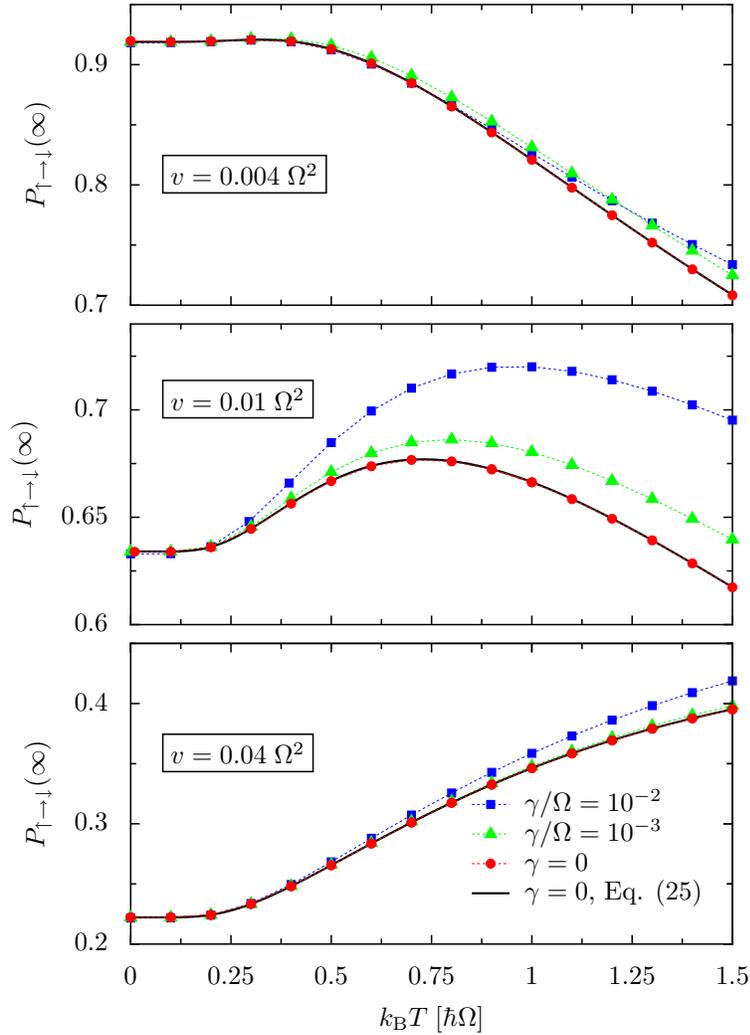}}
\caption{Damping effects on the asymptotic transition probability
$\Pud$ as a function of the temperature for qubit-oscillator coupling
$\g=0.04\Omega$, various sweep velocities $v$ and damping strengths
$\gamma$.
The symbols depict numerical results obtained with the quantum master
equation \eref{QME-cqed-ps}, while the solid lines mark the results
from the independent-crossing approximation \eref{Puuana} for
$\gamma=0$.
}
\label{fig:Tdep}
\end{figure}

We next turn to the generic situation in which both thermal
excitations of the initial state and dissipative transitions play a
role, i.e.\ we consider the situation of finite temperatures and
finite dissipation strength.
The resulting spin-flip probabilities $\Pud$ for three different sweep
velocities are shown in figure~\ref{fig:Tdep}.  For small
temperatures, $\kB T\lesssim 0.2\hbar\Omega$, we find that dissipation
slightly reduces the spin-flip probability $\Pud$.  This is consistent
with the behaviour at zero temperature, discussed in
section~\ref{sec:zeroT}.
Once the thermal energy is of the order $\hbar\Omega$, the opposite
is true: dissipation supports transitions to the final ground state
$|\down\rangle$ and, thus, $\Pud$ increases.  This tendency is most
pronounced for the intermediate sweep velocity chosen in
figure~\ref{fig:Tdep}(b).  In this regime we again find a
non-monotonic temperature dependence of the transition probabilities.

\begin{figure}
\centerline{\includegraphics{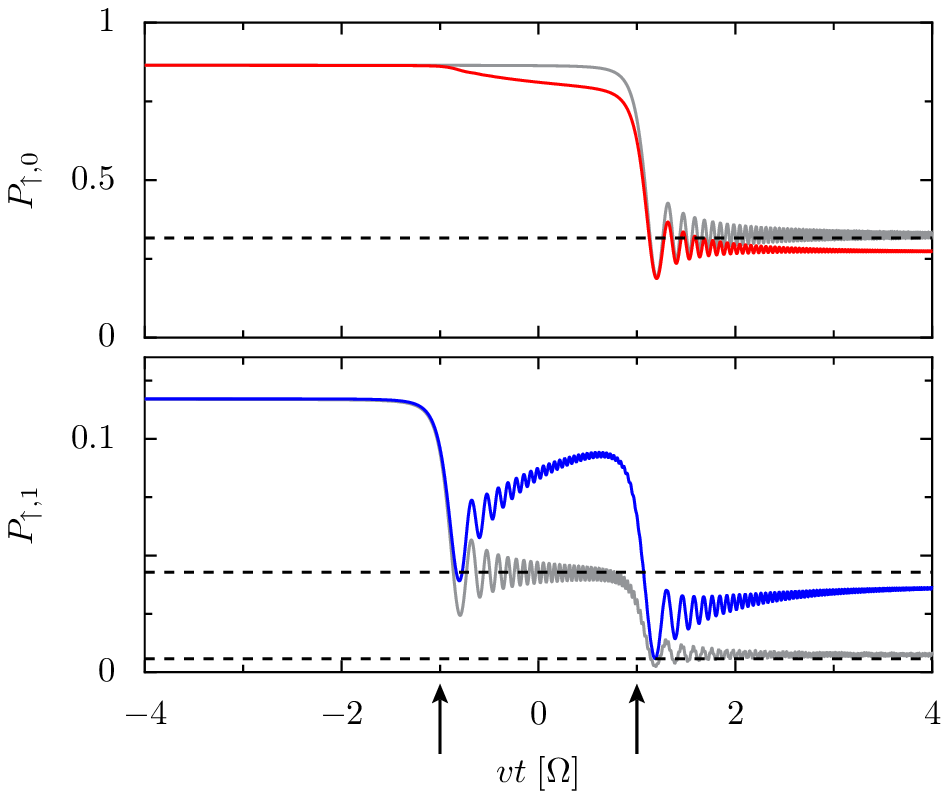}}
\caption{Population dynamics of the states $\ket {\up, 0}$ (a) and
$\ket{\up,1}$ (b) during the Landau-Zener sweep.
Dissipation strength and temperature are
$\dmp=10^{-2}\Omega$ and $\kB T = 0.5\hbar \Omega$, respectively.
The gray lines are the corresponding results in the zero-friction
limit $\gamma=0$, for which the probabilities converge to the values
indicated by the dashed lines.
The arrows mark the positions of the avoided crossings in the qubit
oscillator spectrum; cf.\ figure~\ref{Rabi}.
}
\label{fig:popdyn}
\end{figure}

In order to reveal the role of dissipative decays, we focus on the
population dynamics for an intermediate temperature $\kB T\approx
0.5\hbar\Omega$, where the transition probabilities are already
significantly influenced by thermal excitations; cf.\
figure~\ref{fig:Tdep}.  Nevertheless, the temperature is still
sufficiently low, such that only the oscillator states
$|0\rangle$ and $|1\rangle$ are relevant.  Figure~\ref{fig:popdyn} shows
the time evolution of the population of the states $|\up,0\rangle$ and
$|\up,1\rangle$.  Obviously, the populations change considerably at
the avoided crossings of the qubit-oscillator spectrum, so that the
dynamics discerns into three stages.

For $t<-\Omega/v$, the system remains in the canonical state.  When at
time $t=-\Omega/v$ the anti-crossing between the states
$|\up,1\rangle$ and $|\down,0\rangle$ is reached, the population of the
former state drops due to an adiabatic transition to the latter.
As a consequence, the oscillator is no longer at thermal equilibrium
and, consequently, we observe thermal excitations from $|\up,0\rangle$
to $|\up,1\rangle$.  When at time $t=\Omega/v$, the second set of
anti-crossings is reached (see figure~\ref{fig:paths}(a,b)), both
states undergo an individual Landau-Zener transition
after which the populations converge in an oscillatory manner towards
their final value.  At the final stage, the oscillator populations
thermalize, while those of the qubit remains practically unchanged.
The latter sounds counter-intuitive because the physical system is
dissipative.  Nevertheless this has a physical explanation: The qubit
experiences an effective heat bath with a spectral
density sharply peaked at the oscillator frequency $\Omega$.
Thus for large times, $t\gg\Omega/v$, the spectral density at the
qubit splitting $\hbar vt$ vanishes and, consequently, the qubit is
effectively decoupled from the bath.

The population dynamics provides evidence that for weak dissipation
and narrowly avoided crossings, the dynamics consists of individual
Landau-Zener transitions, while dissipation takes place mainly
in-between the transitions.  This behaviour resembles the one occurring in
nanomagnets \cite{Wernsdorfer1999a, Giraud2001a}.


\section{Conclusions}
\label{sec:c}

We have investigated the influence of finite temperature, decoherence,
and dissipation on Landau-Zener transitions of a two-level system
(qubit) that is coupled via a harmonic oscillator to a heat bath.  In
particular, we have focussed on a recent solid-state realization of
this model, namely the so-called circuit QED for which Landau-Zener
sweeps can be performed by switching the effective Josephson energy of
the Cooper pair box.
The adiabatic spectrum of this system consists of a sequence of exact
and avoided crossings, where for strong qubit-oscillator coupling, the
latter may even overlap.  Therefore, the resulting quantum dynamics is
more complex than in the ``standard'' two-level Landau-Zener problem.
Moreover, this qubit-oscillator-bath model is equivalent to coupling
the qubit to a bath with peaked spectral density.

Dissipation has been modelled by coupling the oscillator to an Ohmic
environment which we integrated out within a Bloch-Redfield approach.
We solved the resulting master equation numerically after a
transformation to Wigner representation followed by a decomposition
into proper basis functions.  The comparison with results for the
exactly solvable zero-temperature limit, demonstrated that our approach
provides reliable results, even though this limit is known to be
rather challenging for a Markovian master equation.

For vanishing dissipation strength, the temperature enters only via
initial thermal excitations of the oscillator.  Most interestingly, we
found for this case that the spin-flip probability exhibits a
non-monotonic temperature dependence.  For a sufficiently small
qubit-oscillator coupling, this can be understood within the
approximation of individual Landau-Zener crossings.  This picture
reveals the special role played by the first excited oscillator state.
As compared to any other state, this state is more likely to induce a
spin-flip.  At intermediate temperatures, the first excited oscillator
state possesses a relatively high influence, which leads to the observed
non-monotonic behaviour.
When dissipative decays become relevant as well, transitions to the
final (adiabatic) ground state of the qubit become more likely.
Nevertheless, in some small regions of parameter space, we find the
opposite, namely that the probability to find the qubit in the excited
state is slightly enhanced.  This is at first sight counter-intuitive,
but can be understood from the fact that the qubit effectively
experiences a bath with a peaked spectral density.  Therefore,
dissipative qubit transitions can occur only during the short lapses of
time in which the adiabatic energy splitting of the qubit is of the
order of the oscillator frequency.  With an increasing oscillator-bath
coupling, the peak in the spectral density becomes broader and, thus,
the more intuitive tendency towards the final ground state starts to
dominate.

Finally, our results provide evidence that the recently derived
zero-temperature results hold true also for finite temperatures,
provided that the thermal energy does not exceed a value of roughly 20
percent of the oscillator's energy quantum.  This means that
experimental quantum state preparation schemes that rely on
Landau-Zener transitions in the zero-temperature limit, are feasible
already when the oscillator is initially in its ground state, while
the low-frequency modes of the bath may nevertheless be thermally
excited.

\section*{Acknowledgements}

It is a pleasure to thank Martijn Wubs and Gert-Ludwig Ingold for
interesting discussions.
We gratefully acknowledge financial support by the German Excellence
Initiative via the ``Nanosystems Initiative Munich (NIM)''.
This work has been supported by Deutsche Forschungsgemeinschaft
through SFB 484 and SFB 631.

\appendix

\section{Derivation of the quantum master equation}
\label{app:qme}

In this Appendix we outline the derivation of the master
equation~\eref{QME-cqed} starting from the general Bloch-Redfield
expression~\eref{BR}.

\subsection{Heisenberg coupling operator}

The essential step is the solution of the
Heisenberg equation of motion for the scaled position operator $\X = a
+ a^{\dag}$ of the oscillator, which will rely on approximations.
In doing so, we even address a slightly more general qubit Hamiltonian
outside the charge degeneracy point, i.e.\ we also consider the
charging energy $\frac{1}{2}\EC\sigma_{x}$, such that the Rabi Hamiltonian
(\ref{Rabi}) becomes
\begin{equation}
\label{rabi-moregral}
\Hs
= - \frac{\EJ}{2} \sigma_{z} - \frac{\EC}{2} \sigma_{x}
  + \hbar \Omega a^{\dag} a + \hbar \g \sigma_{x} (a^{\dag}+a).
\end{equation}
For convenience, we write this Hamiltonian in the eigenbasis of the
qubit:
\begin{equation}
\label{rabi-moregral2}
\Hs = - \frac{\hbar \spl}{2} \bar\sigma_{z}
-  \hbar \Omega a^{\dag} a
+ \hbar\g (\cos\theta \bar\sigma_{z}
          -\sin \theta \bar\sigma_{x} ) (a^{\dag}+a) ,
\end{equation}
where
\begin{equation}
\hbar\spl = \sqrt{\EJ^{2} + \EC^{2}},
\qquad
\theta = \arctan\frac{\EJ}{\EC}
\end{equation}
denote the energy splitting and the coupling angle, respectively, of
the qubit.  The corresponding Heisenberg equations become
\begin{eqnarray}
\label{Qddot}
\ddot \X
&=& -\Omega^{2}\X -g\Omega (\cos\theta \bar\sigma_{z} - \sin\theta \bar\sigma_{x} ),
\\
\dot{\bar\sigma}_{z}
&=& -2 g \sin\theta \bar\sigma_{y} \X ,
\\
\dot{\bar\sigma}_{x}
&=&
(\spl - 2 g \cos\theta) \bar\sigma_{y} ,
\\
\dot{\bar\sigma}_{y}
&=&
-(\spl - 2 g \cos\theta) \bar\sigma_{x} + 2 g \sin\theta \bar\sigma_{z}x,
\end{eqnarray}
which are non-linear due to the qubit-oscillator coupling and, thus,
cannot be solved directly.  We are only interested in the time
evolution of the coordinate $\X$ of the oscillator with eigenfrequency
$\Omega$. Since in typical circuit QED experiments $\g\ll\Omega$, we
treat the coupling as a perturbation.  In the absence of the coupling,
the time evolution of the qubit operators reads
\begin{eqnarray}
\bar\sigma_{x}(t) &=& \bar\sigma_{x} \cos(\spl t) + \bar\sigma_{y} \sin(\spl t) ,
\\
\bar\sigma_{z}(t) &=& \bar\sigma_{z} .
\end{eqnarray}
Inserting this into the equation of motion \eref{Qddot} for the
oscillator coordinate, it becomes evident that the qubit entails on
the oscillator the time-dependent force
\begin{equation}
F(t)
= -\g\Omega \Big\{ \bar\sigma_{z}\cos\theta -
       \big [\bar\sigma_{x}\cos(\spl t) +\bar\sigma_{y}\sin(\spl t)
       \big]\sin\theta \Big \}.
\end{equation}
To first order in $g$, the solution of equation~\eref{Qddot} reads
\begin{equation}
\X (t)
= a \e^{\iu \Omega t} + a^{\dag}  \e^{-\iu \Omega t}
  + \int_{0}^{t} \diff t' \, G(t-t') F(t') ,
\end{equation}
where
\begin{equation}
G(t) = \frac{\sin(\Omega t)}{\Omega} \theta(t)
\end{equation}
denotes the retarded Green function of the classical dissipative
harmonic oscillator.  Evaluating the integral we finally obtain the
Heisenberg operator
\begin{eqnarray}
\nonumber
\X(t)
&=&
a \e^{\iu \Omega t} + a^{\dag}  \e^{-\iu \Omega t}
- \g \Ic(0,\Omega;t) \sigma_{z} \cos\theta
\\
&& + \g \big[ \Ic(\Omega, \spl; t) \sigma_{x}
             +\Is(\Omega, \spl; t) \sigma_{y} \big] \sin\theta ,
\label{Xt-app}
\end{eqnarray}
with the functions
\begin{eqnarray}
\Ic (a,b; t) &=& \frac {a\cos(at) - a\cos(b t)}{a^{2} - b^{2}} ,
\\
\Is (a,b; t)
 &=& \frac {b \sin(a t) - a \sin(b t)}{a^{2} - b^{2}} .
\end{eqnarray}
Expression (\ref{Xt-app}) allows one to explicitly evaluate the
Bloch-Redfield equation \eref{BR}.

\subsection{Ohmic spectral density}

In circuit QED, the environment of the qubit and the transmission line
is formed by electric circuits and, thus, can be characterised by an
effective impedance.  In most cases, this impedance is dominated by an
Ohmic resistor, which corresponds to the Ohmic spectral density
\eref{Jw} of the bath.  Then the time integration in the
Bloch-Redfield equation \eref{BR} can be evaluated.
The antisymmetric correlation function \eref{corr-ant} is then given by
\begin{equation}
\Ant(\tau)
= \frac{\gamma}{2\pi\Omega} \int_0^\infty \diff\tau\, \omega\sin(\omega\tau)
= -\frac{\gamma}{2\Omega}\frac{\diff}{\diff\tau}\delta(\tau) ,
\end{equation}
such that the last term of equation~\eref{BR} becomes $(\iu\gamma/4\Omega) \dot Q$.
The remaining time integrals are of the type
\begin{eqnarray}
\int_{0}^{\infty} \diff \tau \; \Sym (\tau) \cos(\Omega \tau)
&=&
\frac {\Omega}{2} \coth \Big( \frac{\hbar\Omega}{2\kB T} \Big)
\\
\int_{0}^{\infty} \diff \tau \; \Sym (\tau) \sin(\Omega \tau)
&=&
\frac{\nu_{1}  \Omega \cutoff^{2}}{\cutoff^{2} + \Omega^{2}}
\sum_{n=-\infty}^{\infty} \frac{\Omega^{2} - \nu_{n}
\cutoff}{(\nu_{n}+ \cutoff)(\nu_{n}^{2}+ \Omega^{2})} \;,
\end{eqnarray}
where we introduced the Matsubara frequencies $\nu_{n} = 2\pi n \kB T/\hbar$.
In the second integral, an ultraviolet divergence has been regularised
by a Drude cutoff $\e^{-\omega/\cutoff}$ \cite{Weiss1999a}.

\section{Basis expansion}
\label{app:be}

In this appendix we outline the diagonalisation of the oscillator
Liouvillian in Wigner representation, $\Lho$, whose eigenvectors are
used as a basis set for the numerical treatment.  Since the operator
$\Lho$, apart from the cross-diffusion $D_{xp}$, is of the same form as
the Fokker-Planck operator of the corresponding problem for classical
Brownian motion, we can make use of an idea put forward by Titulaer
\cite{Titulaer1978a, Risken1989a} and generalise it along the lines of
Ref.~\cite{Kohler1997a}.

\subsection{Diagonalisation of the oscillator Liouvillian}
\label{app:Lho}

By solving the characteristic functions of the partial differential
equation $\dot\phi=\Lho\phi$, one finds the operators
\begin{eqnarray}
\label{transfuno}
\rai  &=& \partial_{x} + \frac{\eig}{\Omega} \partial_{p} \, ,
\\
\label{transfdos}
\low &=&
\frac{\Omega^{2}}{\eig^{2}-\Omega^{2}}
\Big ( \Diff \partial_{x} - \frac{\eig}{\Omega}  \Dpp \partial_{p} + x
      - \frac{\eig}{\Omega}  p \Big) ,
\end{eqnarray}
which commute with $\partial/\partial t-\Lho$ and, thus, map any
solution of the Liouville equation to a further solution.
For notational convenience, we have introduced the eigenvalues of the
classical equation of motion of the dissipative harmonic oscillator,
\begin{equation}
\eig = -\frac{\gamma}{2} + \iu \sqrt{\Omega^{2} - \frac{\gamma^{2}}{4}} \; ,
\end{equation}
and $\lambda^*$.  It will turn out that the equilibrium expectation
value of the dimensionless oscillator coordinate $Q$ becomes $ \Diff =
\Dpp + \dmp \Dxp/\Omega $.  Since moreover, the diffusion coefficient
$\Dxp$ appears in all results only in this combination, the
introduction of $\Diff$ turns out to be convenient as well; see also
the discussion after equation \eref{Fsigma}.  The operators
\eref{transfuno} and \eref{transfdos} fulfil the commutation relations
\begin{eqnarray}
[\low, \rai ] = 1 \; , \quad
[\low, \rai^* ] = [\low^*,\rai] = 0 \; , \quad
\left [ \Lho, \railow \right]  = \pm \eig \railow
\end{eqnarray}
and allow one to write the Liouvillian in the diagonal form
\begin{equation}
\Lho = \eig \rai \low + \eigc \raic \lowc \; ,
\end{equation}
where the symbol $^*$ denotes complex conjugation.\footnote{We do not
consider the overdamped limit $\dmp >2$ $\eig$ in which $\eig$ becomes
real and, thus, the notation needs to be modified.}
Formally we have reduced the eigenvalue problem for the Liouvillian to
that of two uncoupled harmonic oscillators, so that the eigenvalues
obviously read
\begin{equation}
\lambda_{nn'}
= \eig n + \eigc n'
= -(n+n')\frac{\gamma}{2} +\iu(n-n')\sqrt{\Omega^2-\frac{\gamma^2}{4}} \,,
\end{equation}
where $n,n'=0,1,2,\ldots$
The corresponding eigenfunctions $\eif_{nn'}$ can be constructed by
applying the raising operators $\rai$ and $\raic$ on the ``ground
state''.  The latter is the stationary state defined by the relation
$\Lho\eif_{00} = 0$ whose solution is the Gaussian
\begin{equation}
\eif_{00}(x,p)
=
\frac{\Omega}{2 \pi  \sqrt{ \langle \X^{2} \rangle_{{\rm eq}}
         \langle \dot \X^{2} \rangle_{{\rm eq}} }}
\exp\Big(
 -\frac{\Omega^{2} p^{2}}{2  \langle \dot \X^{2} \rangle_{{\rm eq}}}
 -\frac{x^{2}}{2 \langle Q^{2} \rangle_{{\rm eq}} }
\Big) .
\end{equation}

Since $\Lho \neq \Lho^{\dag}$, the eigenfunctions $\eif_{nn'}$ are not
mutually orthogonal, so that we need to compute the left eigenvectors
$\leif_{nn'}$ of $\Lho$ as well.  Repeating the calculation from above for
$\Lho^\dag$, we find the left ground state $\leif_{00}=1$, so that we
obtain the eigenfunctions
\begin{eqnarray}
\eif_{nn'}
&=&
\frac{1}{n! n' !}
(\rai)^{n} (\raic)^{n'} \eif_{00}
\\
\leif_{nn'} &=& (\lrai)^{n} (\lraic)^{n'} \; 1
\end{eqnarray}
which fulfil the ortho-normalisation relation
\begin{equation}
\int \diff x \diff p \, \leif_{mm'} \eif_{nn'} = \delta_{nm} \delta_{n'm'} .
\end{equation}

\subsection{Expansion of the entire Liouvillian}

Besides the already diagonalized oscillator Liouvillian, the total
Liouvillian \eref{QME-cqed-ps} for the qubit plus the oscillator
contains also the operators
\begin{eqnarray}
\label{xtrans}
x &=& - \low - \lowc
    + \sima\rai+\sima^*\raic , \qquad
    \sima = \Diff\frac{\Omega ^{2}}{\eig^{2} -\Omega^{2}},
\\
\label{dptrans}
\partial_{p}
&=& -\iu \frac{\rai-\raic}{2 \sqrt{1-\dmp^{2}/4 \Omega^{2}}} .
\end{eqnarray}
Then, the basis decomposition of the Wigner function,
\begin{equation}
\label{expansion}
\Wm = \sum_{nn'} \cmatrixnn  \eif_{nn'} \; ;
\qquad
\cmatrixnn
=
\left ( \begin{array}{cc} \cnn ^{\up \up} & \cnn^{\up \down} \\
                          \cnn^{\down \up} & \cnn^{\down \down}
        \end{array} \right) ,
\end{equation}
obeys the equation of motion
\begin{eqnarray}
\nonumber
\dot \cmatrixnn
&=&
  (n\eig+n'\lambda^*) \cmatrixnn
  + \iu \frac{\EJ}{2 \hbar} [\sigma_{z}, \cmatrixnn]
  -\iu  \g  [\sigma_{x},  \cmatrixnpn + \cmatrixnnp]
\\ && \nonumber
+ \iu \frac{\g}{\sqrt{1 - \dmp^{2}/4\Omega^{2}}}
  [\sigma_{x},  n' \cmatrixnnm - n \cmatrixnmn ]_{+}
\\ && \nonumber
  + \iu \frac{\g}{\sqrt{1 - \dmp^{2}/4\Omega^{2}}}
  [\sigma_{x}, \sima  n \cmatrixnmn - \sima^{*} n' \cmatrixnnm]
\\ &&
  -\iu \frac{\gamma\Dsigma}{\sqrt{1 - \dmp^{2}/4\Omega^{2}}}
   [\sigma_{x},  n' \cmatrixnnm - n \cmatrixnmn ] ,
\label{QME-coeff}
\end{eqnarray}
which is a set of $4N^2$ coupled linear ordinary differential equations.

\subsection{Computation of expectation values}

The expectation value of an operator can be performed directly in the
basis of the eigenfunctions without back transformation to the operator
representation of the density operator.  For an observable
\begin{equation}
\Ob = \Q \otimes \Osc(x,p) ,
\end{equation}
for which $S$ and $\Osc$ refer to qubit and oscillator variables,
respectively, the expectation value
\begin{equation}
\langle \Ob \rangle
=
\tr \big(\Ob \dm \big)
=
\sum_{ij} \Q_{ij} \int \diff x \diff p  \; \Osc (x, p)
\sum_{nn'} \cnn^{ij} \eif_{nn'}(x,p)
\end{equation}
can be expressed in terms of the operators \eref{transfuno} and
\eref{transfdos} such that
\begin{equation}
\langle \Ob \rangle
= \sum_{ij} \sum_{nn'} \Q_{ij} \Osc_{nn'} \cnn^{ij}(t) \, ,
\end{equation}
where $\Osc (x, p)$ is the corresponding operator in phase-space
\cite{Hillery1984a} and
\begin{equation}
\Osc_{nn'} = \int \diff x \diff p\, O(x,p)
\eif_{nn'}(x,p).
\end{equation}
If one is only interested in the behaviour of the qubit, i.e.\
for $O=1$, the expectation value becomes
\begin{equation}
\langle B\rangle = \langle \Q \rangle = \sum_{ij}  \Q_{ij} \;  c_{00}^{ij}(t) .
\end{equation}
This implies that any information about the qubit state is already
contained in the four coefficients $c_{00}^{ij}(t)$, which
represents a particular advantage in the present decomposition.

\section*{References}
\bibliographystyle{iopart}
\bibliography{decoherence}

\end{document}